\documentclass[prd,superscriptaddress,amsfonts,amssymb,amsmath,showpacs,twocolumn]{revtex4-2}
\usepackage{bm}
\usepackage{amsfonts}
\usepackage{latexsym}
\usepackage[latin1]{inputenc}
\usepackage{graphicx}
\usepackage{amsmath}
\usepackage{palatino}
\usepackage{ragged2e}
\usepackage{mathpazo}
\usepackage{textcomp}
\usepackage{float}
\usepackage{booktabs}
\usepackage{dcolumn}
\usepackage{multirow}
\usepackage{hyperref}
\hypersetup{colorlinks,citecolor=blue}
\usepackage{amsmath}
\usepackage{xcolor}
\usepackage{orcidlink}
\usepackage[caption=false]{subfig}
\usepackage{commath}
\def\jnl@style{\it}
\def\aaref@jnl#1{{\jnl@style#1}}

\def\aaref@jnl#1{{\jnl@style#1}}

\def\aj{\aaref@jnl{AJ}}                   
\def\apj{\aaref@jnl{ApJ}}                 
\def\apjl{\aaref@jnl{ApJ}}                
\def\apjs{\aaref@jnl{ApJS}}               
\def\apss{\aaref@jnl{Ap\&SS}}             
\def\aap{\aaref@jnl{A\&A}}                
\def\aapr{\aaref@jnl{A\&A~Rev.}}          
\def\aaps{\aaref@jnl{A\&AS}}              
\def\mnras{\aaref@jnl{Mon.~Not.~Roy.~Astron.~Soc.}}             
\def\prd{\aaref@jnl{Phys.~Rev.~D}}        
\def\prc{\aaref@jnl{Phys.~Rev.~C}}  
\def\prl{\aaref@jnl{Phys.~Rev.~Lett.}}    
\def\qjras{\aaref@jnl{QJRAS}}             
\def\skytel{\aaref@jnl{S\&T}}             
\def\ssr{\aaref@jnl{Space~Sci.~Rev.}}     
\def\zap{\aaref@jnl{ZAp}}                 
\def\nat{\aaref@jnl{Nature}}              
\def\aplett{\aaref@jnl{Astrophys.~Lett.}} 
\def\apspr{\aaref@jnl{Astrophys.~Space~Phys.~Res.}} 
\def\physrep{\aaref@jnl{Phys.~Rep.}}      
\def\physscr{\aaref@jnl{Phys.~Scr}}       
\def\commat{\aaref@jnl{Comm.~Math.~Phys.}}              
\def\science{\aaref@jnl{Science}}               
\def\cqg{\aaref@jnl{Classical Quant.~Grav.}}            
\def\jpcs{\aaref@jnl{JPCS}}                                     
\def\ijmpd{\aaref@jnl{Int.~J.~Mod.~Phys.~D}}                    
\def\grg{\aaref@jnl{Gen.~Relat.~Gravit.}}               
\def\rpp{\aaref@jnl{Rep.~Prog.~Phys.}}          
\def\npa{\aaref@jnl{Nucl.~Phys.~A}}        
\def\lrr{\aaref@jnl{Living Rev.~Rel.}}                   
\def\jcap{\aaref@jnl{J.~Cosmology Astropart.~Phys.}}    
\def\rmp{\aaref@jnl{Rev.~Mod.~Phys.}}   
\def\epjc{\aaref@jnl{Eur.~Phys.~J.~C}}

\allowdisplaybreaks[1]

\begin{document}

\color{black}       

\title{Wormhole solutions under the effect of dark matter in $f(R,L_m)$ gravity}

\author{Lakhan V. Jaybhaye\orcidlink{0000-0003-1497-276X}}
\email{lakhanjaybhaye@gmail.com}
\affiliation{Department of Mathematics, Birla Institute of Technology and
Science-Pilani,\\ Hyderabad Campus, Hyderabad-500078, India.}
\author{Moreshwar Tayde\orcidlink{0000-0002-3110-3411}}
\email{moreshwartayde@gmail.com}
\affiliation{Department of Mathematics, Birla Institute of Technology and
Science-Pilani,\\ Hyderabad Campus, Hyderabad-500078, India.}
\author{P.K. Sahoo\orcidlink{0000-0003-2130-8832}}
\email{pksahoo@hyderabad.bits-pilani.ac.in}
\affiliation{Department of Mathematics, Birla Institute of Technology and
Science-Pilani,\\ Hyderabad Campus, Hyderabad-500078, India.}

\date{\today}

\begin{abstract}

In the background of $f(R, L_m)$ gravity, this work investigates three distinct dark matter halo profiles to test the possibility of generalised wormhole geometry within the galactic halo regions. The current study aims to accomplish these goals by examining various dark matter profiles including Universal Rotation Curves (URC), Navarro-Frenk-White (NFW) model-I, and NFW model-II inside two distinct $f(R, L_m)$ gravity models. According to the $f(R, L_m) = \frac{R}{2} + L_m^\alpha$ model, the DM halo density profiles produce suitable shape functions that meet all the necessary requirements for exhibiting the wormhole geometries with appropriate choice of free parameters. In addition, to examine DM profiles under the $f(R, L_m) = \frac{R}{2} + (1 + \lambda R)L_m$ model, we consider a specific shape function. Further, we observed that the derived solution from both two models violates the null energy constraints, confirming that the DM supports wormholes to maintain in the galactic halo.
\end{abstract}

\maketitle

\section{Introduction}\label{sec1}

A wormhole is a solution derived from Einstein's field equations that offers a potential means of connecting different regions of spacetime or even different universes \cite{Ein}. The concept of a wormhole was initially proposed by L. Flamm \cite{Fla}. The renowned Einstein-Rosen bridge (ERB), which was first defined by Einstein and Rosen in 1935, is where the idea of wormholes first emerged. The search for traversable wormholes has been an interesting area of study since it was discovered that the initial wormhole solution was not feasible \cite{Whe,Ful,Ell}. A multitude of general relativistic models of traversable wormholes have been put proposed. For example, Ellis suggested that a traversable wormhole might be supported by a phantom scalar field \cite{Ell}. Thorne presented the well-known Morris-Thorne (MT) traversable wormhole solution \cite{Mor1}, which depends on exotic matter to keep the wormhole's throat open. As exotic matter has a certain mix of negative pressure and positive energy density, it creates a repulsive force that keeps the wormhole from collapsing. Typically, such exotic matter violates conventional energy conditions \cite{Vis}. Notably, traversable wormholes were created by Kanti et al. in the framework of quadratic gravitational theories, in which the wormhole throat remains open due to gravity processes alone, eliminating the need for exotic matter \cite{Kan,Kan1}.\\ 
Currently, the most widely accepted theory to explain gravity is General Relativity (GR), which is backed by a large number of experiments and observations. However some occurrences are still beyond the capacity of GR to fully explain. These consist of the accelerated expansion of the universe that has been observed, the unexpected effects of gravity on galactic shapes, and the search for a quantum framework that can explain gravity. Other theories of gravity have been put forth in response to these issues. GR is extended or modified in different ways by these theories. The known $f(R)$ gravity is one example of how GR can be substantially extended.  According to this theory, a function of the scalar curvature takes the place of the Ricci scalar in the Einstein-Hilbert action \cite{Sta,Buc,Fel}. This modification affects the behaviour of gravity at various scales and results in modifications to the gravitational field equations.\\
Many astrophysicists have recently expressed an interest in the study of wormhole geometry. In reference \cite{Y. Ahmad}, Zubair et al. conducted an examination of static spherically symmetric wormhole geometry, exploring various configurations with anisotropic, isotropic, and barotropic matter content. Moreover, using barotropic and anisotropic equations of state cases, wormhole solutions have been studied in symmetric teleparallel gravity \cite{Tay3} and found the violation of Null Energy Condition (NEC). Additionally, Mustafa et al. \cite{Zuba2} investigated wormhole solutions that contravene the NEC within the framework of Rastall gravity. Also, in geometry-matter couplings \cite{Sah}, it is shown that the matter content of the wormhole remarkably able to obey the energy conditions. There are a number of modified gravity theories that have been proposed in the literature \cite{Mus2,Has,Eli,Tay,Tay1,Tay2,Zuba,Shar3,Wahe,Azma,Zuba1,Zuba3,Mus3} to get around the undetectable exotic matter. \\
An enhancement of the $f(R)$ modified gravity framework involves introducing a direct connection between the arbitrary function of the Ricci curvature $R$, and the matter Lagrangian term $L_m$, as put forth in \cite{Ber}. This instance was subsequently extended to arbitrary matter-geometry couplings by Harko and Lobo \cite{Har}. There are several important astrophysical and cosmological uses for cosmological models with non-minimal curvature-matter couplings \cite{Har1,Har2,Har3,Har4,Far}. $f(R, L_m)$ modified gravity, which integrates curvature-matter coupling theories, was recently presented by Harko and Lobo \cite{Har5}. Here, $f(R, L_m)$ is a generic function of the Lagrangian term $L_m$ and the Ricci curvature $R$. Within the $f(R, L_m)$ modified gravity framework, the covariant divergence of the energy-momentum tensor doesn't vanishes, leading to an additional force orthogonal to the four velocities. Consequently, the motion of a test particle follows a non-geodesic path. Furthermore, cosmological models incorporating $f(R, L_m)$ gravity do not adhere to the equivalence principle, which has been constrained by experimental tests within the solar system \cite{Far1,Ber1}. Notably, there have been several intriguing cosmological and astrophysical studies conducted in the field of $f(R, L_m)$ gravity theory, with references \cite{Gon,Lab1,Lab2,Har6,Har7,Sol} serving as valuable sources for further exploration.\\
Dark matter (DM) is a basic element of the Universe that is mostly observable via its gravitational effect, not its luminosity. In order to first suggest that DM exists in galaxies, Zwicky \cite{Zwi} employed the virial theorem. URC is thought to be present in spiral galaxies, and the gravitational effect of DM in their galactic halos is indicative of its presence \cite{Rob,Ros}. Based on observations evidence, including the NFW density profile and the flat rotation curves observed in galaxies, Rahaman et al. \cite{Far2} illustrated the feasibility of traversable wormholes finding support within galactic halos. Subsequently, the exploration of galactic halo wormholes has become a subject of interest in various gravitational theories, including modifications. For example, Sharif et al. \cite{Sha,Sha1,Sha2} investigated solutions involving galactic halo wormholes within different modified gravity theories. Furthermore, this area has seen additional compelling research contributions from others \cite{Mus,Mus1,Ovg,Xu,Kuh}. In this work, we explore traversable wormhole solutions in the context of the $f(R, L_m)$ theory of gravity by employing DM halo profiles. In order to accomplish this, we examine different DM density profile models, specifically the URC model and the cold dark matter halo with two different NFW models. This work attempts to prove that, for the models under consideration, traversable wormholes in the galactic halo are really present in the context of $f(R, L_m)$ gravity.\\
This work is organized as follows: Section \ref{sec2} lays out the fundamental formulations of the $f(R, L_m)$ gravity theory. Section \ref{sec3} focuses on deriving the field equations for the static and spherically symmetric Morris-Thorne wormhole metric using a general $f(R, L_m)$ function. Section \ref{sec4} delves into a specific $f(R, L_m)$ model, namely, $f(R, L_m) = \frac{R}{2} + L_m^\alpha$, which is applied to describe DM halo profiles. Section \ref{sec5} explores another non-linear $f(R, L_m)$ model, specifically, $f(R, L_m) = \frac{R}{2} + (1 + \lambda R)L_m$, and examines its implications for DM halo profiles, along with the associated shape functions, $b(r)=\text{\text{e}} ^{1-\frac{r}{r_0}} \cosh \left(\frac{r}{r_0}\right)$. Section \ref{sec6} provides a summary of the study's findings and initiates a discussion on the results obtained throughout the research.

\section{ $f(R,L_m)$ Gravity Theory}\label{sec2}
We provide a quick introduction to $f(R,L_m)$  gravity in this section. Harko et al.\cite{Har5} provide the action for $f(R,L_m)$  gravity as
\begin{equation}\label{1a}
S= \int{f(R,L_m)\sqrt{-g}\text{d}^4x}\,.
\end{equation}
The Ricci scalar, denoted as $R$ , is associated with the metric tensor $g_{\mu\nu}$, which has a determinant represented as $g$, while $L_m$ signifies the matter Lagrangian. The Ricci scalar curvature term $R$ can be derived through the contraction of the Ricci tensor $R_{\mu\nu}$ as
\begin{equation}\label{1b}
R= g^{\mu\nu} R_{\mu\nu}\,,
\end{equation} 
where,
\begin{equation}\label{1c}
R_{\mu\nu}= \partial_\lambda \Gamma^\lambda_{\mu\nu} - \partial_\nu \Gamma^\lambda_{\lambda\mu} + \Gamma^\sigma_{\mu\nu} \Gamma^\lambda_{\sigma\lambda} - \Gamma^\lambda_{\nu\sigma} \Gamma^\sigma_{\mu\lambda}\,,
\end{equation}
where $\Gamma^\alpha_{\beta\gamma}$  represents the components of the Levi-Civita connection, which can be computed as
\begin{equation}\label{1d}
\Gamma^\alpha_{\beta\gamma}= \frac{1}{2} g^{\alpha\lambda} \left( \frac{\partial g_{\gamma\lambda}}{\partial x^\beta} + \frac{\partial g_{\lambda\beta}}{\partial x^\gamma} - \frac{\partial g_{\beta\gamma}}{\partial x^\lambda} \right).
\end{equation}
The field equation obtained through the variation of the general action \eqref{1a} associated with the metric tensor $g_{\mu\nu}$ is as follows:
\begin{multline}\label{1e}
f_R R_{\mu\nu} + (g_{\mu\nu} \square - \nabla_\mu \nabla_\nu)f_R - \frac{1}{2} (f-f_{L_m}L_m)g_{\mu\nu}\\
=\frac{1}{2} f_{L_m} T_{\mu\nu}.
\end{multline}
In this context, $f_R$ is defined as the partial derivative of $f$ with respect to $R$, $f_{L_m}$ is defined as the partial derivative of $f$ with respect to $L_m$, and $T_{\mu\nu}$ represents the energy-momentum tensor for the cosmic fluid, which is expressed as follows:
\begin{equation}\label{1f}
T_{\mu\nu} = \frac{-2}{\sqrt{-g}} \frac{\delta(\sqrt{-g}L_m)}{\delta g^{\mu\nu}}.
\end{equation}
Furthermore, the contraction of the field equation \eqref{1e} results in the following relationship between the energy-momentum scalar $T$, the Lagrangian term $L_m$, and the Ricci scalar $R$
 \begin{equation}\label{1g}
R f_R + 3\square f_R - 2(f-f_{L_m}L_m) = \frac{1}{2} f_{L_m} T,
\end{equation}
where $\square F$ represents the d'Alembertian operator applied to a scalar function $F$,\\
defined as $\square F = \frac{1}{\sqrt{-g}} \partial_\alpha (\sqrt{-g} g^{\alpha\beta} \partial_\beta F)$.

\section{Wormhole Geometries in $f(R,L_m)$ Gravity}\label{sec3}
Facilitating a connection between two distant sectors of the universe, wormholes offer a fascinating prospect. The presence of dark matter becomes evident through the flat, rotating trajectories exhibited by neutral hydrogen clouds in the outer regions of spiral galaxies. Within these galactic systems, the outer neutral hydrogen clouds are treated as test particles, following circular orbits that can be accurately described within the framework of static, and spherically symmetric spacetime. Morris and Thorne \cite{Mor} derived a metric for traversable wormholes that exhibits these characteristics, providing a static and spherically symmetric framework as follows:
\begin{equation}\label{3a}
\hspace{-0.3cm} \text{d}s^2=-\text{e}^{2\Phi(r)}\text{d}t^2+\left(1-\frac{b(r)}{r}\right)^{-1}\text{d}r^2+r^2\text{d}\theta^2+r^2 \text{sin}^2\theta \text{d}\phi^2\,.
\end{equation} 
The shape of a wormhole is determined by a function called $b(r)$. This function captures the essence of the structure of wormhole. Additionally, $\Phi(r)$ is the redshift function linked to the gravitational redshift phenomenon. For a wormhole to be traversable, the shape function $b(r)$ must adhere to a crucial condition known as flaring-out condition, expressed as $(b-b'r)/b^2>0$ \cite{Mor}. This means that the shape of wormhole must gracefully open up. At the wormhole throat, marked as $r_0$, the condition $b(r_0)=r_0$ is required, and the derivative, $b^{\,\prime}(r_0)$ should be less than 1. Furthermore, the asymptotic flatness condition is essential, it states that as we move to distant regions $r\rightarrow \infty$, the ratio  $\frac{b(r)}{r}$ should tend to zero. In addition, to ensure there is no event horizon, the redshift function $\Phi(r)$ must remain finite at all points. In the context of Einstein's GR, satisfying these criteria might hint at the presence of exotic matter at the wormhole throat. Furthermore, the appropriate radial distance $l(r)$, which can be expressed as
\begin{equation}\label{3b}
l(r)=\pm \int_{r_0}^{r}\frac{\text{d}r}{\sqrt{1-\frac{b(r)}{r}}}\,.
\end{equation}
This quantity needs to remain finite throughout. In this context, the $\pm$ symbols denote the upper and lower segments of the wormhole, which are interconnected through the throat. Additionally, the proper distance diminishes as we move from the upper universe, where $l=+\infty$, down to the throat, and then further from $l=0$ to $-\infty$ in the lower universe. Furthermore, $l$ should be equal to or greater than the coordinate distance, ensuring that $\mid l(r)\mid \geq r-r_0$.\\
Next, we calculated the Ricci curvature scalar $R$ for the spherically symmetric configuration \eqref{3a} using equation \eqref{1b}:
\begin{multline}\label{3c}
R=\frac{2b'}{r^2} - 2\left( 1-\frac{b}{r} \right) \left\lbrace \Phi''+\Phi'^2+\frac{\Phi'}{r} \right\rbrace \\
+ \frac{\Phi'}{r^2} \left( rb'+b-2r \right).
\end{multline}
In this study, we examine wormhole solutions while taking into account an anisotropic energy-momentum tensor. The expression for this tensor, as outlined in \cite{Kuh}, is defined as
\begin{equation}\label{3d}
T_{\mu\,\nu}=\left(\rho+p_t\right)u_{\mu}\,u_{\nu}+p_t\,\delta_{\mu\,\nu}+\left(p_r-p_t\right)v_{\mu}\,v_{\nu}\,.
\end{equation}
In this context, $u_{\mu}$ represents the four-velocity vector, and $v_{\mu}$ is the unitary space-like vector, both satisfying the conditions $-u_{\mu}u^{\nu}=v_{\mu}v^{\nu}=1$. The energy density is denoted as $\rho$, while $p_r$ and $p_t$ represent the radial and tangential pressures, respectively, with their dependence solely on the radial coordinate $r$. \\
Now, when we incorporate the metric \eqref{3a} and the anisotropic fluid \eqref{3d} into the equations of motion \eqref{1e}, we derive the following field equations
\begin{multline}\label{3e}
\left( 1-\frac{b}{r} \right) \left[ \left\lbrace  \Phi''+\Phi'^2  + \frac{2\Phi'}{r} - \frac{(rb'-b)}{2r(r-b)}\Phi' \right\rbrace F -\left\lbrace \Phi' \right.\right.\\\left.\left.
+\frac{2}{r}- \frac{(rb'-b)}{2r(r-b)}  \right\rbrace F' - F'' \right] + \frac{1}{2} \left( f-L_m f_{L_m} \right) = \frac{1}{2} f_{L_m} \rho\,,
\end{multline}
\begin{multline}\label{3f}
\left( 1-\frac{b}{r} \right) \left[ \left\lbrace  - \Phi''-\Phi'^2  + \frac{(rb'-b)}{2r(r-b)} \left( \Phi' +\frac{2}{r} \right) \right\rbrace F  \right.\\\left.
+ \left\lbrace \Phi'+\frac{2}{r}- \frac{(rb'-b)}{2r(r-b)}  \right\rbrace F' \right] - \frac{1}{2} \left( f-L_m f_{L_m} \right) = \frac{1}{2} f_{L_m} p_r\,,
\end{multline}
\begin{multline}\label{3g}
\left( 1-\frac{b}{r} \right) \left[ \left\lbrace - \frac{\Phi'}{r}  + \frac{(rb'+b)}{2r^2(r-b)} \right\rbrace F+ \left\lbrace \Phi'+\frac{2}{r} \right.\right.\\\left.\left.
- \frac{(rb'-b)}{2r(r-b)}  \right\rbrace F' + F'' \right] - \frac{1}{2} \left( f-L_m f_{L_m} \right) = \frac{1}{2} f_{L_m} p_t\,,
\end{multline}
where $F=\frac{\partial f}{\partial R}$.
By utilizing these particular field equations, it opens the door to the exploration of various wormhole solutions in the framework  of $f(R,L_m)$ gravity models. This opens up the way for studying a diverse array of wormhole configurations and characteristics within the domain of $f(R,L_m)$ gravity.\\
Let's now take a moment to discuss the classical energy conditions, which are derived from the Raychaudhuri equations. These conditions are essential for exploring the physically plausible configurations of matter. Below, you'll find an expression for each of the four energy conditions: the null energy condition (NEC), weak energy condition (WEC), dominant energy condition (DEC), and strong energy condition (SEC).\\
$\bullet$ Weak energy condition (\textbf{WEC}): $\rho\geq0$,\,\, $\rho+p_r\geq0$,\,\, and \,\, $\rho+p_t\geq0$.\\
$\bullet$ Null energy condition (\textbf{NEC}): $\rho+p_r\geq0$,\,\, and \,\, $\rho+p_t\geq0$.\\
$\bullet$ Dominant energy condition (\textbf{\textbf{DEC}}): $\rho\geq0$,\,\, $\rho+p_r\geq0$,\,\, $\rho+p_t\geq0$,\,\, $\rho-p_r\geq0$,\,\, and \,\, $\rho-p_t\geq0$.\\
$\bullet$ Strong energy condition (\textbf{SEC}): $\rho+p_r\geq0$,\,\, $\rho+p_t\geq0$,\,\, and \,\, $\rho+p_r+2p_t\geq0$.\\
In summary, energy conditions are significant restrictions on the behavior of matter in the Universe and play a crucial role in the study of wormholes.
\section{Wormhole Solutions for specific $f(R,L_m)$ Model-I}\label{sec4}
The wormhole solutions in this section are obtained by assuming the following minimal $f(R,L_m)$ function \cite{Jay}
\begin{equation}\label{4a} 
f(R,L_m)=\frac{R}{2}+L_m^\alpha\,.
\end{equation}
In the context of this model, the parameter $\alpha$ holds a significant role as a free and unconstrained variable, allowing for its adjustment to influence the characteristics of system. Notably, when $\alpha$ assumes the specific value of 1, we seamlessly recover the familiar and well-established wormhole geometry as described within the framework of GR.\\
Hence, we can derive the field equations associated with this particular $f(R,L_m)$ function, which are represented by Eq. \eqref{4a}, and when we set $L_m=\rho$ with a constant redshift function, the field equations \eqref{3e}-\eqref{3g} can be simplified as
\begin{equation}\label{4b}
\rho=\left(\frac{b'}{ (2\alpha-1)r^2}\right)^\frac{1}{\alpha}\,,
\end{equation}
\begin{equation}\label{4c}
p_r=\frac{\rho\left( (\alpha-1)r^3 -b \rho^\alpha \right)}{\alpha r^3}\,,
\end{equation}
\begin{equation}\label{4d}
p_t=\frac{\rho^{(1-\alpha)}\left( b-r b' +2(\alpha-1) r^3  \rho^\alpha \right)}{2\alpha r^3}\,.
\end{equation}
Now, in the following subsections, we will explore wormhole solutions under two DM halo profiles namely URC and NFW.

\subsection{URC Model}
This subsection studies the energy density profile of the URC model, which is described by the following equation \cite{Sal}
\begin{equation}\label{4e}
\rho =\frac{\rho_0 r_s ^3}{(r+r_s)(r^2+r_s ^2)}\,.
\end{equation}
Here, we introduce the symbols, $r_s$ denotes the characteristic radius, and $\rho_0$ signifies the central density of the URC dark matter halo. Now, when we compare equations \eqref{4b} and \eqref{4e}, we arrive at the differential equation that characterizes the URC model
\begin{equation}\label{4e1}
 b'= (2\alpha-1)r^2\left(\frac{\rho_0 r_s ^3}{(r+r_s)(r^2+r_s ^2)}\right)^\alpha\,.
\end{equation}
In this case, the shape function is obtained by altering $\alpha$. Specifically, we are considering $\alpha=2,\,3$ for this reason.
\subsubsection{$\alpha = 2$}
\begin{figure*}
\centering
\includegraphics[width=19.5cm,height=9cm]{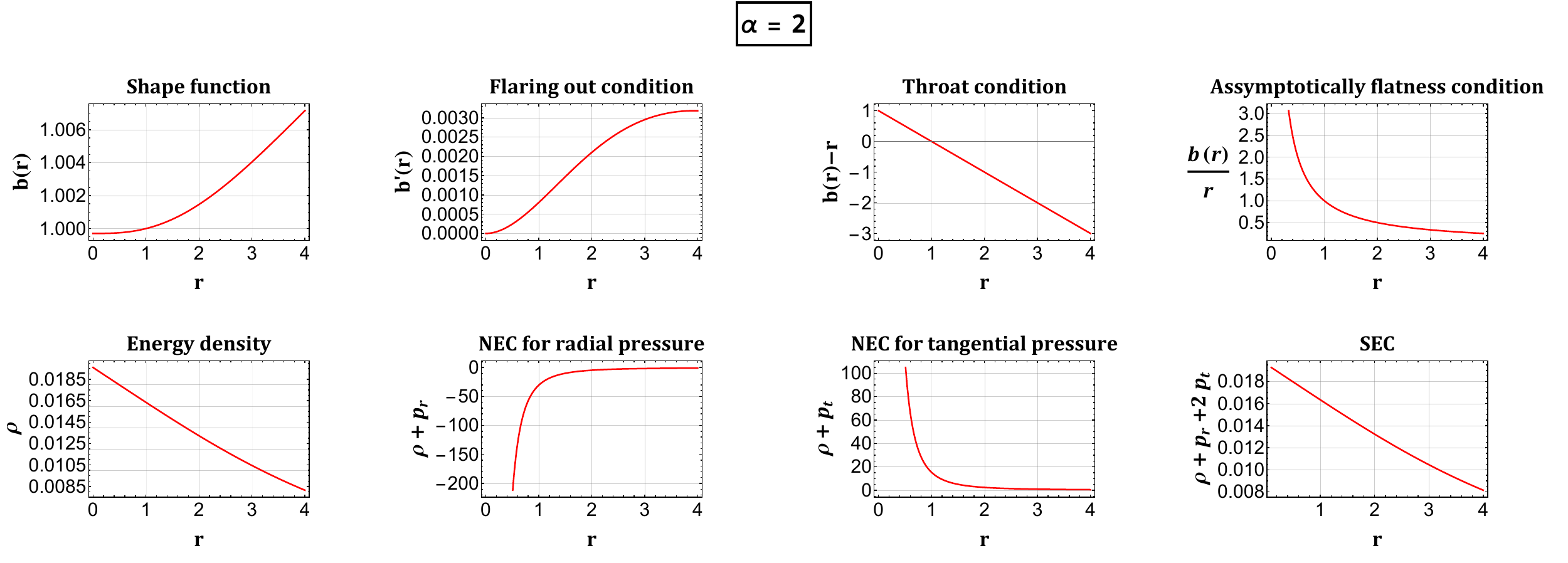}
\caption{Profile shows the behavior of essential conditions of shape function for a wormhole to be traversable and the energy conditions versus radial coordinate $r$ for URC model. Also, we consider $\rho_0=0.0196$, $r_s=6$, and $r_0=1$.}\label{1}
\end{figure*}

 On integrating Eq. \eqref{4e1} for the shape function $b(r)$, we get
\begin{equation}\label{4e2}
b(r)= -\frac{3\rho_0 ^2 r_s ^3}{4}  \left(\frac{r_s \left(2 r_s ^2+r_s r+r^2\right)}{r_s ^3+r_s ^2 r+r_s r^2+r^3}+\tan ^{-1}\left(\frac{r}{r_s}\right)\right)+c_1    \,,
\end{equation}
where the integrating constant is denoted by $c_1$. Using the preceding equation, we now apply the throat condition $b(r_0) = r_0$ to obtain $c_1$ as
\begin{equation}\label{4e3}
 c_1=\frac{3\rho_0 ^2 r_s ^3 }{4} \left(\frac{r_s \left(2 r_s ^2+r_s r_0+r_0 ^2\right)}{r_s ^3+r_s ^2 r_0+r_s r_0 ^2+r_0 ^3}+\tan ^{-1}\left(\frac{r_0}{r_s}\right)\right)+r_0\,,
\end{equation}
where $r_0$ is the throat radius. Substituting the value of $c_1$ in Eq. \eqref{4e2} we get,
\begin{multline}\label{4e4}
 b(r)= r_0 -\frac{3\rho_0 ^2 r_s ^3}{4}  \left[r_s \left(\frac{r_s}{r_s^2+r^2}+\frac{1}{r_s+r}\right)+\tan ^{-1}\left(\frac{r}{r_s}\right) \right.\\\left.
  -r_s \left(\frac{r_s}{r_s^2+r_0 ^2}+\frac{1}{r_s+r_0}\right)-\tan ^{-1}\left(\frac{r_0}{r_s}\right)\right]\,.
\end{multline}
Further, its graphical representation is shown in the Fig. \ref{1}.
Now, the incorporation of Eq. \eqref{4e4} into Eqs. \eqref{4c}-\eqref{4e} for the NEC regarding radial and tangential pressures at the throat $r = r_0 $ can be uniquely expressed as
\begin{multline}\label{4e5}
\rho + p_r \bigg\vert_{r=r_0}=\sqrt{\rho_0^2 r_s^5\mathcal{M}_1}\left(\frac{3}{2}-\frac{1}{2 r_0^2 \rho_0^2 r_s^5 \mathcal{M}_1}\right)\,,
\end{multline}
\begin{multline}\label{4e6}
\rho + p_t \bigg\vert_{r=r_0} =\sqrt{\rho_0^2 r_s^5\mathcal{M}_1}\left(\frac{3}{4}+\frac{1}{4 r_0^2 \rho_0^2 r_s^5 \mathcal{M}_1}\right)\,,
\end{multline}
where $\mathcal{M}_1=\frac{{r_s} }{({r_0}+{r_s})^2 \left({r_0}^2+{r_s}^2\right)^2}$\,.
\subsubsection{$\alpha = 3$}
\begin{figure*}
\centering
\includegraphics[width=19.5cm,height=9cm]{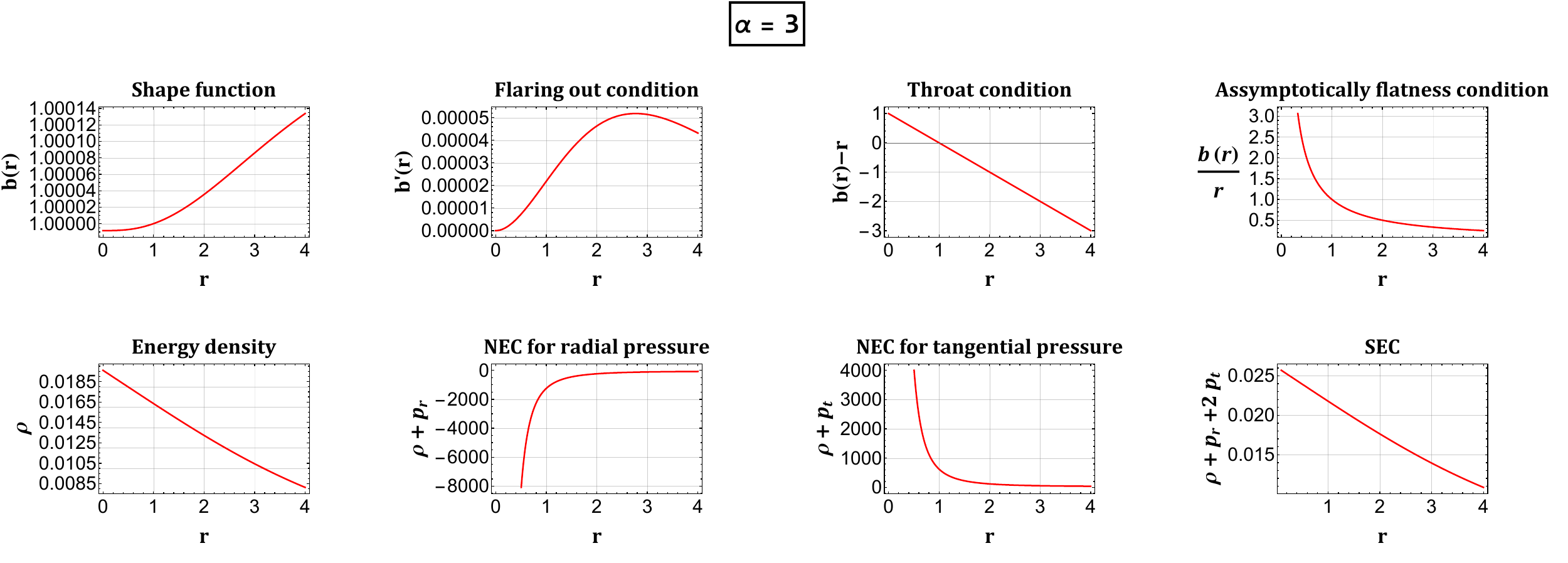}
\caption{Profile shows the behavior of essential conditions of shape function for a wormhole to be traversable and the energy conditions versus radial coordinate $r$ for URC model. Also, we consider $\rho_0=0.0196$, $r_s=6$, and $r_0=1$.}\label{2}
\end{figure*}

By performing another integration on Eq. \eqref{4e1} with $\alpha=3$ to determine the shape function $b(r)$, using the throat condition $b(r_0)=r_0$, we arrive at the following result

\begin{multline}\label{4f2}
 b(r)= r_0 -\frac{5\rho_0^3 {r_s}^3}{32}  \left[ \mathcal{M}_2\left(16 r^3 {r_s}^2+20 r^2 {r_s}^3+12 r^4 {r_s}+5 
\right.\right. \\ \left.\left.
\hspace{1cm}r^5+15 r {r_s}^4+12 {r_s}^5\right)-\log \left(r^2+{r_s}^2\right)+2 \log (r+{r_s})
\right.  \\ \left.
\hspace{1cm}+7 \tan ^{-1}\left(\frac{r}{{r_s}}\right) -\mathcal{M}_1\left(16 {r_0}^3 {r_s}^2 +20 {r_0}^2 {r_s}^3 +12 {r_0}^4 
\right.\right. \\ \left.\left.
\hspace{1cm}{r_s}+5 {r_0}^5+15 {r_0} {r_s}^4+12 {r_s}^5\right)+\log \left({r_0}^2+{r_s}^2\right)-2
 \right. \\ \left.
 \log ({r_0}+{r_s})
 -7 \tan ^{-1}\left(\frac{{r_0}}{{r_s}}\right)\right]\,.
\end{multline}

where $\mathcal{M}_2=\frac{{r_s} }{({r_s}+r)^2 \left({r_s}^2 +r^2\right)^2}$.\\
Also, its graphical representation is shown in the Fig. \ref{2}.
Now, to study the energy conditions, the incorporation of Eq. \eqref{4f2} into Eqs. \eqref{4c}-\eqref{4e} for the NEC regarding radial and tangential pressures at the throat $r = r_0 $ can be uniquely expressed as
\begin{equation}\label{4f3}
\rho + p_r \bigg\vert_{r=r_0}=(\mathcal{M}_3)^{1/3}\left(\frac{5}{3}-\frac{1}{3 r_0^2 \mathcal{M}_3}\right)\,,
\end{equation}
\begin{equation}\label{4f4}
\rho + p_t \bigg\vert_{r=r_0} =(\mathcal{M}_3)^{1/3}\left(\frac{5}{6}+\frac{1}{6 r_0^2 \mathcal{M}_3}\right)\,,
\end{equation}
where $\mathcal{M}_3=\frac{\rho_0^3 r_s^9}{(r_0+r_s)^3 \left(r_0^2+r_s^2\right)^3}$.\\
One can explore the behavior of NEC at the throat in the Fig. \ref{2}.
\subsection{NFW Model-I}
Hernquist \cite{Her} introduced a potential density, with a primary focus on investigating both theoretical and observational facets pertaining to elliptical galaxies. Following this, Navarro and colleagues \cite{Nav} conducted an analysis of equilibrium density profiles within dark matter (DM) halos in universes characterized by hierarchical clustering. Their investigation, employing high-resolution N-body simulations, revealed a consistent profile shape across various parameters, including halo mass, spectral shape of initial density fluctuations, and values of the cosmological parameter.\\
Now, we start by taking a look at the energy density distribution with one of the NFW models, which is defined as 
\begin{equation}\label{4g}
\rho =\frac{\rho_0 r_s}{r \left(\frac{r}{r_s}+1\right)^2}\,.
\end{equation}
Here, $\rho_0$ signifies the central density of the NFW Model-I dark matter halo. Now, when we compare equations \eqref{4b} and \eqref{4g}, we arrive at the differential equation that characterizes the NFW Model-I as
\begin{equation}\label{4g1}
 b'= (2\alpha-1)r^2\left(\frac{\rho_0 r_s}{r(\frac{r}{r_s}+1) ^2}\right)^\alpha\,.
\end{equation}
In this case also, the shape function is obtained by altering $\alpha$. Specifically, we are considering $\alpha=2,\,3$ for this reason. 
\subsubsection{$\alpha = 2$}
\begin{figure*}
\centering
\includegraphics[width=19.5cm,height=9cm]{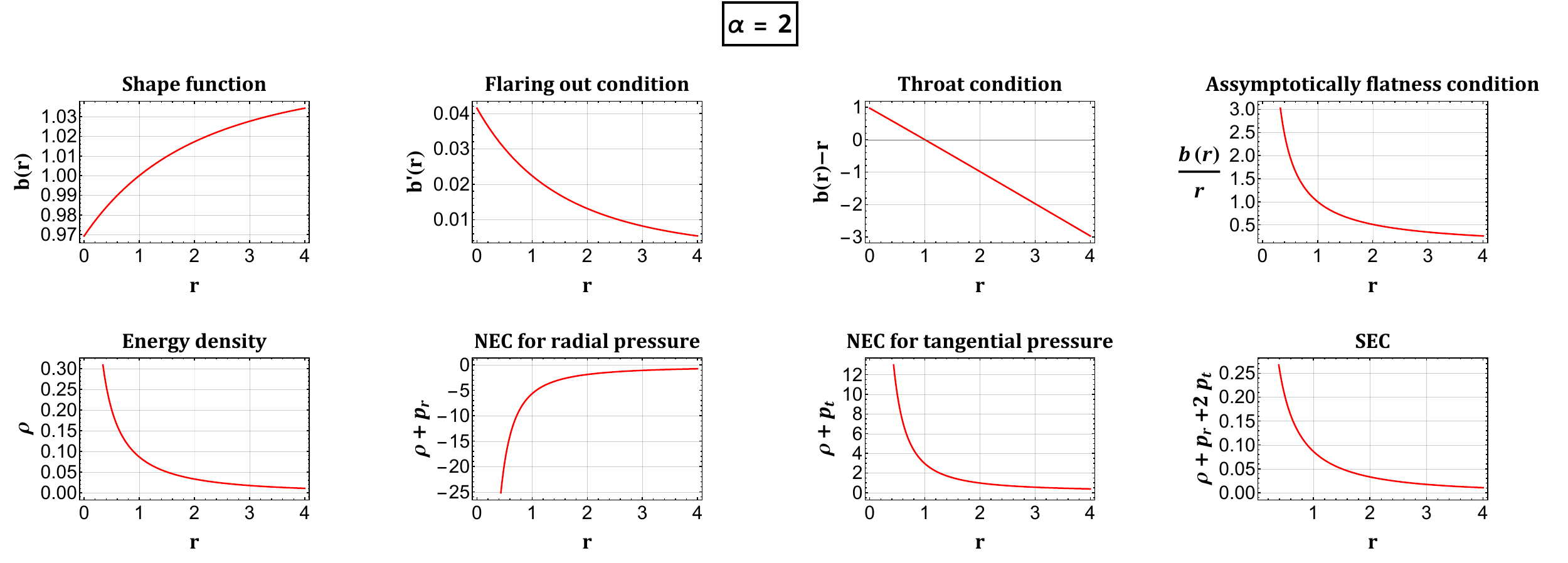}
\caption{Profile shows the behavior of essential conditions of shape function for a wormhole to be traversable and the energy conditions versus radial coordinate $r$ for NFW-1 model. Also, we consider $\rho_0=0.0196$, $r_s=6$, and $r_0=1$.}\label{3}
\end{figure*}

On integrating Eq. \eqref{4g1} for the shape function $b(r)$, we get

\begin{equation}\label{4g4}
 b(r)= r_0-\frac{\rho_0 ^2 r_s ^6}{(r_s+r)^3}+\frac{\rho_0 ^2 r_s ^6}{(r_s+r_0)^3}\,.
\end{equation}
Further, its graphical representation is shown in Fig. \ref{3}.
Now, the NEC for radial and tangential pressures at the throat $r = r_0$ using Eq. \eqref{4g4} into Eqs. \eqref{4c}, \eqref{4d} and \eqref{4g} can be read as
\begin{equation}\label{4g5}
 \rho+p_r\bigg\vert_{r=r_0}=\frac{2 r_0 ^2 \mathcal{M}_4 ^3 +r_0 ^2 \mathcal{M}_4 ^2 -1}{2 r_0 ^2 \mathcal{M}_4} \,,
\end{equation}
\begin{equation}\label{4g6}
 \rho+p_t\bigg\vert_{r=r_0}=\frac{4 r_0 ^2 \mathcal{M}_4 ^3 - r_0 ^2 \mathcal{M}_4 ^2 +1}{4 r_0 ^2 \mathcal{M}_4} \,,
\end{equation}
where $\mathcal{M}_4=\frac{{\rho _0  r_s ^3} }{r_0  (r_s +r_0 ^2 )^2}.$
\subsubsection{$\alpha = 3$}
Now for $\alpha=3$, integrating Eq. \eqref{4g1} for the shape function $b(r)$, we get

\begin{multline}\label{4h2}
 b(r)= r_0+\frac{\rho_0 ^3 r_s ^3}{12}  \left[\mathcal{M}_5 \left(137 r_s ^4+385 r_s ^3 r+470 r_s ^2 r^2 +270
 \right.\right. \\ \left.\left.
 \hspace{1cm}  r_s r^3+60 r^4\right)-60 \log (r_s+r)+60 \log (r)-\mathcal{M}_6
  \right. \\ \left.
 \hspace{1cm} \left(137 r_s ^4+385 r_s ^3 r_0+470 r_s ^2 r_0 ^2+270 r_s r_0 ^3 +60 r_0 ^4\right) 
 \right. \\ \left.
-60 \log (r_s+r_0)+60 \log (r_0)\right]\,,
\end{multline}
where $\mathcal{M}_5=\frac{{ r_s} }{ (r_s + r)^5}$, and $\mathcal{M}_6=\frac{{ r_s} }{ (r_s +r_0 )^5}$.\\

\begin{figure*}
\centering
\includegraphics[width=19.5cm,height=9cm]{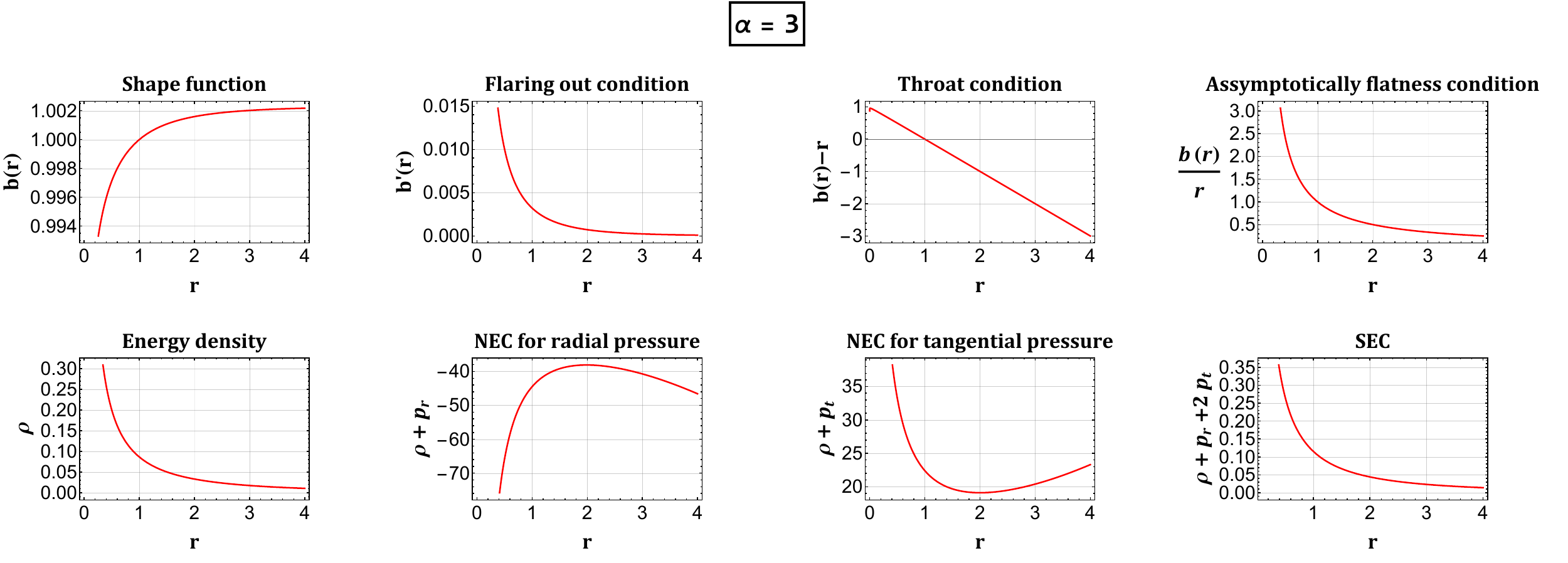}
\caption{Profile shows the behavior of essential conditions of shape function for a wormhole to be traversable and the energy conditions versus radial coordinate $r$ for NFW-1 model. Also, we consider $\rho_0=0.0196$, $r_s=6$, and $r_0=1$.}\label{4}
\end{figure*}

\justify The NEC for radial and tangential pressures at the throat $r = r_0$ using Eq. \eqref{4h2} into Eqs. \eqref{4c}, \eqref{4d} and \eqref{4g} can be read as
\begin{multline}\label{4h3}
\rho+p_r\bigg\vert_{r=r_0}=\frac{(\mathcal{M}_7) ^{1/3}}{3 \rho_0 ^3 r_s ^9} \left(2 r_0 ^3 r_s ^9+3 \rho_0  r_s ^3 r_0 ^2 (r_s+r_0)^4 
\right. \\ \left.
(\mathcal{M}_7) ^{2/3}-r_0(r_s+r_0)^6\right)\,,
\end{multline}
\begin{equation}\label{4h4}
\rho+p_t\bigg\vert_{r=r_0}=\frac{\rho_0 r_s ^3}{r_0 (r_s+r_0)^2} +\frac{r_0 -\mathcal{M}_7}{6 r_0 ^3 (\mathcal{M}_7)^{2/3}}\,,
\end{equation}
where $\mathcal{M}_7=\frac{{\rho _0 ^3 r_s ^9} }{r_0 ^3 (r_s +r_0 )^6}$.\\
One can check the behavior of NEC at the throat $r=r_0$ from Fig. \ref{4}.
\subsection{NFW Model-II}
Within this subsection, we will again consider another form of energy density for NFW Model-II, which can be expressed using the following equation
\begin{equation}\label{4i}
\rho =\frac{\rho_0 r_s}{r \left(\frac{r}{r_s}+1\right)^3}\,,
\end{equation}
where $\rho_0$ indicates the central density of NFW Model-II dark matter halo. Upon comparing Equations \eqref{4b} and \eqref{4i}, the differential equation under NFW Model-II can be obtained as 
 \begin{equation}\label{4i1}
 b'= (2\alpha-1)r^2\left(\frac{\rho_0 r_s}{r \left(\frac{r}{r_s}+1\right)^3}\right)^\alpha\,.
\end{equation}
To obtain the shape function, we are changing $\alpha$ in this case also. We are specifically taking into consideration $\alpha=2,\,3$ for this purpose.  
\subsubsection{$\alpha = 2$}
On integrating Eq. \eqref{4i1} with $\alpha=2$ for the shape function $b(r)$, using the same procedure used in URC and NFW-1 models, we get

\begin{equation}\label{4i4}
 b(r)= r_0 -\frac{3 \rho_0 ^2 r_s ^8}{5 (r_s+r)^5}+\frac{3 \rho_0 ^2 r_s ^8}{5 (r_s+r_0)^5}\,.
\end{equation}
For the NEC, the reduction of Eq. \eqref{4i4} into Eqs. \eqref{4c}, \eqref{4d} and \eqref{4i} about the radial and tangential pressures at the throat $r = r_0$ can be uniquely stated as 
\begin{equation}\label{4i5}
\rho + p_r \bigg\vert_{r=r_0}=\frac{3\mathcal{M}_8 ^2 -1}{2 r_0 \mathcal{M}_8}\,,
\end{equation}
\begin{equation}\label{4i6}
\rho + p_t \bigg\vert_{r=r_0} =\frac{1-3\mathcal{M}_8 ^2}{4 r_0 \mathcal{M}_8}\,,
\end{equation}
where $\mathcal{M}_8=\frac{\rho_0 r_s ^4 }{(r_0+r_s)^3} $.
\begin{figure*}
\centering
\includegraphics[width=19.5cm,height=9cm]{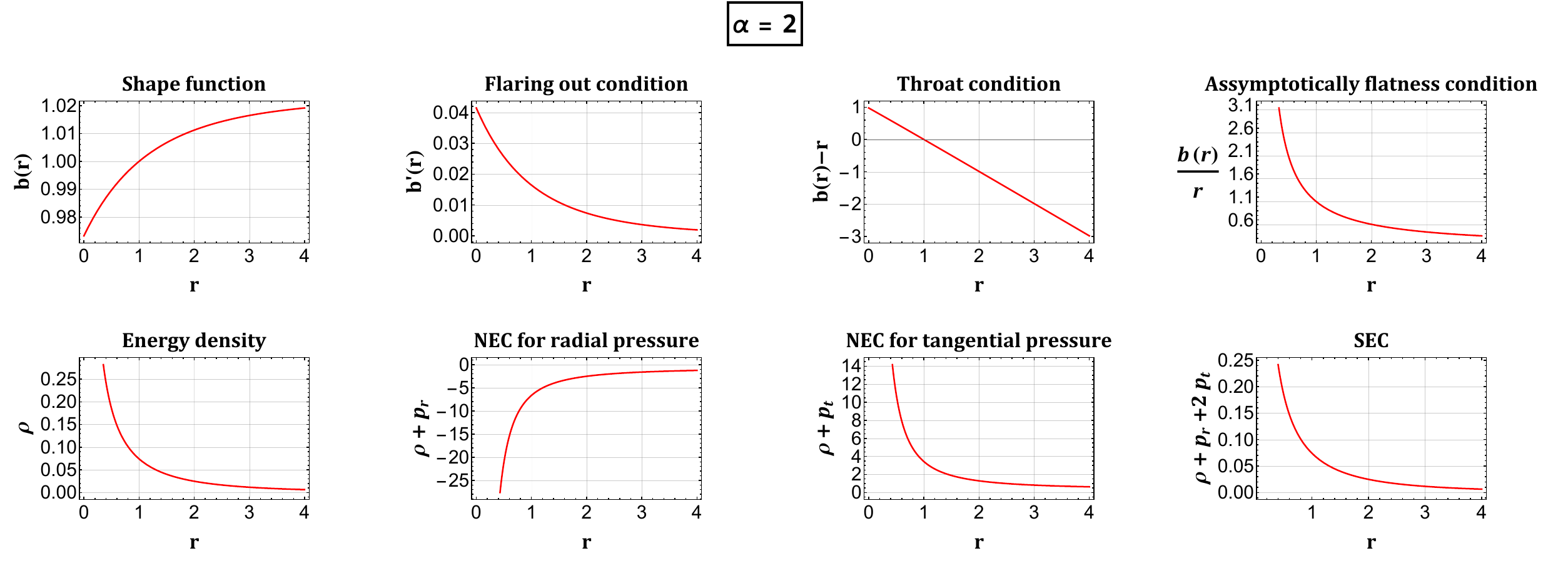}
\caption{Profile shows the behavior of essential conditions of shape function for a wormhole to be traversable and the energy conditions versus radial coordinate $r$ for the NFW-II model. Also, we consider $\rho_0=0.0196$, $r_s=6$, and $r_0=1$.}\label{5}
\end{figure*}
\subsubsection{$\alpha = 3$}
Now again integrating Eq. \eqref{4i1} with $\alpha=3$ for the shape function $b(r)$, we obtain

\begin{multline}\label{4j2}
 b(r)=r_0+ \frac{\rho_0 ^3 r_s ^3}{168}  \left[\mathcal{M}_9\left(2283 r_s ^7+11544 r_s ^6 r+28644 r_s ^5 r^2
\right.\right. \\ \left.\left.
\hspace{1cm}+41608 r_s ^4 r^3+37310 r_s ^3 r^4+20440 r_s ^2 r^5+6300 r_s r^6 +
\right.\right. \\ \left.\left.
\hspace{1cm}840 r^7\right)-840 \log (r_s+r)+840 \log (r)- \mathcal{M}_{10}\left(2283
\right.\right. \\ \left.\left.
\hspace{1cm} r_s ^7+11544 r_s ^6 r_0+28644 r_s ^5 r_0 ^2+41608 r_s ^4 r_0 ^3+37310 r_s ^3 
\right.\right. \\ \left.\left.
\hspace{1cm}r_0 ^4 +20440 r_s ^2 r_0 ^5+6300 r_s r_0 ^6+840 r_0 ^7\right)-840 \log (r_s
\right. \\ \left.
+r_0)+840 \log (r_0)\right],
\end{multline}\,
where $\mathcal{M}_9=\frac{ r_s } { (r_s +r )^8}$, and $\mathcal{M}_{10}=\frac{ r_s } { (r_s +r_0 )^8}.$\\

 Using Eq. \eqref{4j2} to enter Eqs. \eqref{4c}, \eqref{4d}, and \eqref{4i}, the NEC for radial and tangential pressures at the throat $r = r_0$ can be interpreted as
\begin{equation}\label{4j3}
\rho + p_r \bigg\vert_{r=r_0}=\frac{5\mathcal{M}_{8} ^3 -r_0}{3 r_0 \mathcal{M}_{8} ^2}\,,
\end{equation}
\begin{equation}\label{4j4}
\rho + p_t \bigg\vert_{r=r_0} =\frac{r_0-5\mathcal{M}_{8} ^3}{6 r_0 \mathcal{M}_{8}^2}\,.
\end{equation}
\begin{figure*}
\centering
\includegraphics[width=19.5cm,height=9cm]{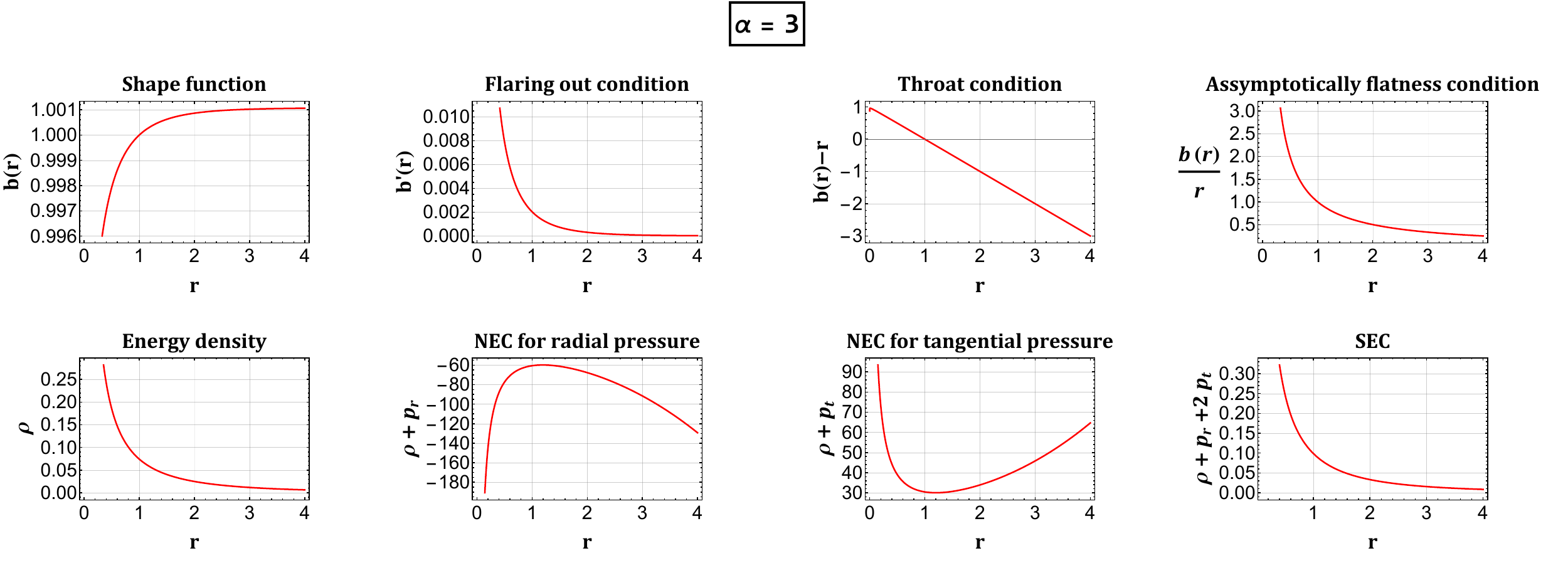}
\caption{Profile shows the behavior of essential conditions of shape function for a wormhole to be traversable and the energy conditions versus radial coordinate $r$ for the NFW-II model. Also, we consider $\rho_0=0.0196$, $r_s=6$, and $r_0=1$.}\label{6}
\end{figure*}
Now, in the next subsection (\ref{subsectionD}), we will discuss the conditions of the shape function for the wormhole to be traversable and the energy conditions.
\subsection{Discussion}\label{subsectionD}
Here, we will discuss the graphical representation of the acquired shape function and the prerequisites for the existence of a wormhole obtained in the URC and NFW profiles as well as the energy conditions.  To accomplish this, we make some choices regarding the relevant model parameter $\alpha$ as $2$ and $3$ to obtain the exact solutions. Also, we fix another free parameter as $\rho_0=0.0196$, $r_s=6$, and $r_0=1$. Initially, our examination focuses on the behavior of shape functions and the prerequisites for the existence of a wormhole, assuming a constant redshift function. In Figs. \ref{1} - \ref{6}, we depict the characteristics of the shape function and the flaring out conditions, all across two different values of $\alpha$. Notably, the shape function shows a favorably increasing behavior. Also, it underscores the satisfaction of the flaring out condition, as denoted by $b'(r_0)<1$, at the wormhole throat. Shifting our attention to the third plot, we consider the wormhole throat at $r_0 = 1$, and its graphical representation is featured in it. In this specific instance, insights into the asymptotic behavior of the shape function across a range of $\alpha$ values. It reveals that, as the radial distance increases, the ratio $\frac{b(r)}{r}$ gradually approaches zero, thus ensuring the asymptotic behavior of the shape function.\\
Further, we studied the energy conditions and their graphical representation is shown in the same Figs. \ref{1} - \ref{6}. It displays the energy density as a function of radial distance ($r$), illustrating a consistent and positive decrease throughout the entire spacetime. In contrast, a negative trend for the radial NEC indicates a violation of the NEC. On the other hand, it demonstrates a positive behavior for the tangential NEC. Finally, the behavior of the SEC for two different values of $\alpha$, shows a consistent positive behavior.
\section{Wormhole Solutions for specific $f(R,L_m)$ Model-II}\label{sec5}
The wormhole solutions in this section are obtained by assuming the following non-minimal $f(R,L_m)$ function \cite{Lab2},
\begin{equation}\label{5a}
f(R,L_m)=\frac{R}{2}+(1+\lambda\,R)L_m\,.
\end{equation}
Here, $\lambda$ represents the coupling constant. Notably, when $\lambda=0$, we rediscover the conventional wormhole geometry of GR. The cosmological implications of this model were previously explored in \cite{Sol}. We will now investigate the features of this particular model in the context of wormholes. Consequently, we derive the following field equations associated with the specific $f(R,L_m)$ function.
\begin{equation}\label{5b}
\rho = \frac{b'}{r^2 +2\lambda b'}\,,
\end{equation}
\begin{equation}\label{5c}
p_r=\frac{2\lambda r(r^3 -b')b'-(r^2 + 4\lambda b') b}{r(r^2 +2\lambda b')^2}\,,
\end{equation}
\begin{equation}\label{5d}
p_t=\frac{r^2 b -(r^3-4\lambda b) b'}{2r(r^2 +2\lambda b')^2}\,.
\end{equation}
The wormhole solution is obtained in this study by using the redshift function $\Phi(r)=\text{constant}$. As it is difficult to obtain the exact solution for the non-linear model, we are considering a particular form of the shape function which is as follows:
\begin{equation}\label{5e}
b(r)=\text{\text{e}} ^{1-\frac{r}{r_0}} \cosh \left(\frac{r}{r_0}\right)\,.
\end{equation}
Now, in the following subsections (\ref{subsectionA}-\ref{subsectionC}), we will study the energy conditions using the shape function \eqref{5e} for the URC and NFW profiles.
\subsection{URC Model}\label{subsectionA}
\begin{figure*}
\centering
\includegraphics[width=18.5cm,height=5.5cm]{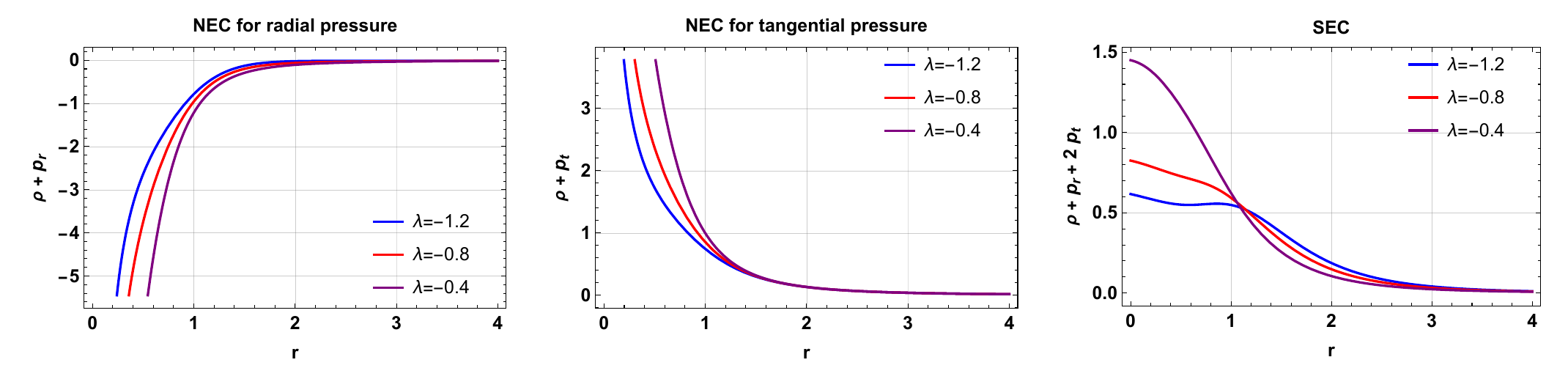}
\caption{Profile shows the behavior of NEC and SEC versus radial coordinate `$r$ ' for different values of model parameter $\lambda$ under URC model. Also, we consider $\rho_0=0.2$, $r_s=1.4$, and $r_0=1.5$.}\label{7}
\end{figure*}

In this subsection, our focus will be on exploring the energy density profile of the URC model, which we previously defined in Eq. \eqref{4e}. To gain a deeper understanding of this model, we will analyze the NEC for both radial and tangential pressures at the throat $r = r_0$, by incorporating the shape function from Eq. \eqref{5e} into Equations \eqref{5c} and \eqref{5d}. This comprehensive examination will provide us with valuable insights into the behavior and properties of this model-II.
\begin{multline}\label{5f}
 \rho+p_r\bigg\vert_{r=r_0}=(\mathcal{M}_3)^{1/3}- \frac{\text{\text{e}} \left(4\lambda  r_0 ^4 +\left(1+\text{\text{e}}^2\right) r_0 ^3-4\lambda  \text{\text{e}} \right)}{2 \left(\text{\text{e}} r_0 ^3-2 \lambda \right)^2}\,,
\end{multline}
\begin{multline}\label{5g}
 \rho+p_t\bigg\vert_{r=r_0}=(\mathcal{M}_3)^{1/3}+ \frac{\text{\text{e}} \left(3+\text{\text{e}}^2\right) r_0 ^3-4 \left(1+\text{\text{e}}^2\right) \lambda }{4 \left(\text{\text{e}} r_0 ^3-2 \lambda \right)^2}\,.
\end{multline}
And, the graphical representation for energy conditions is shown in Fig. \ref{7}.
\subsection{NFW Model-I}\label{subsectionB}
\begin{figure*}
\centering
\includegraphics[width=18.5cm,height=5.5cm]{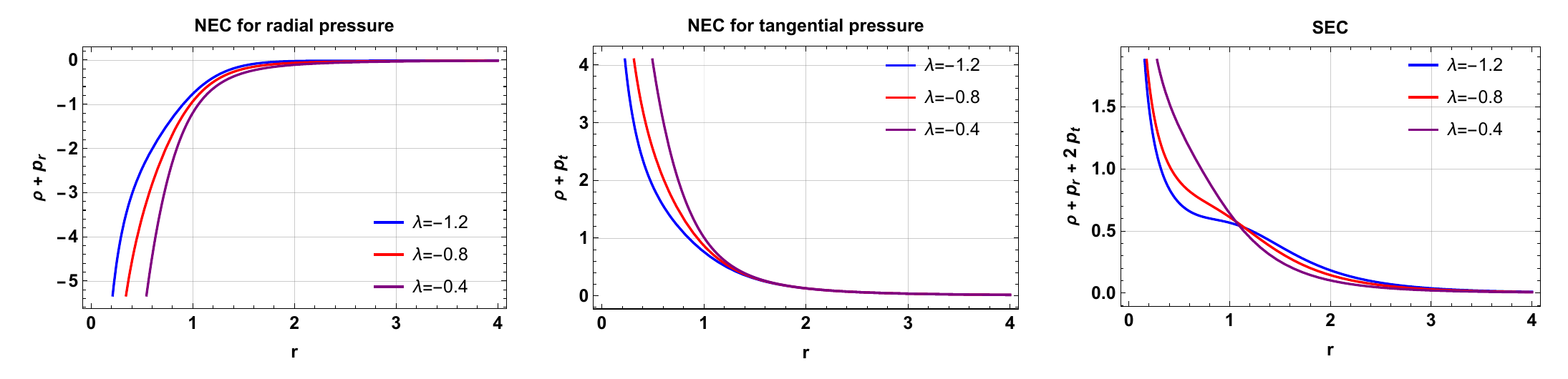}
\caption{Profile shows the behavior of NEC and SEC versus radial coordinate `$r$ ' for different values of model parameter $\lambda$ under NFW-1 model. Also, we consider $\rho_0=0.2$, $r_s=1.4$, and $r_0=1.5$.}\label{8}
\end{figure*}

In this specific subsection, we will investigate the energy density profile of the NFW Model-I, which we previously specified in Eq. \eqref{4g}. By inserting the shape function from Eq. \eqref{5e} into Equations \eqref{5c} and \eqref{5d}, we will analyse the Null Energy Condition for both radial and tangential pressures at the throat $r = r_0$. 
\begin{multline}\label{5h}
\rho+p_r\bigg\vert_{r=r_0}= \mathcal{M}_4 -\frac{\text{\text{e}} \left(4 r_0 ^4 \lambda +\left(1+\text{\text{e}}^2\right) r_0 ^3-4 \text{\text{e}} \lambda \right)}{2 \left(\text{\text{e}} r_0 ^3-2 \lambda \right)^2}\,,
\end{multline}
\begin{multline}\label{5i}
 \rho+p_t\bigg\vert_{r=r_0}=\mathcal{M}_4+\frac{\text{\text{e}} \left(3+\text{\text{e}}^2\right) r_0 ^3-4 \left(1+\text{\text{e}}^2\right) \lambda }{4 \left(\text{\text{e}} r_0 ^3-2 \lambda \right)^2}\,.
\end{multline}
Furthermore, one can observe the graphical depiction of energy conditions in Fig. \ref{8}.
\subsection{NFW Model-II}\label{subsectionC}
\begin{figure*}
\centering
\includegraphics[width=18.5cm,height=5.5cm]{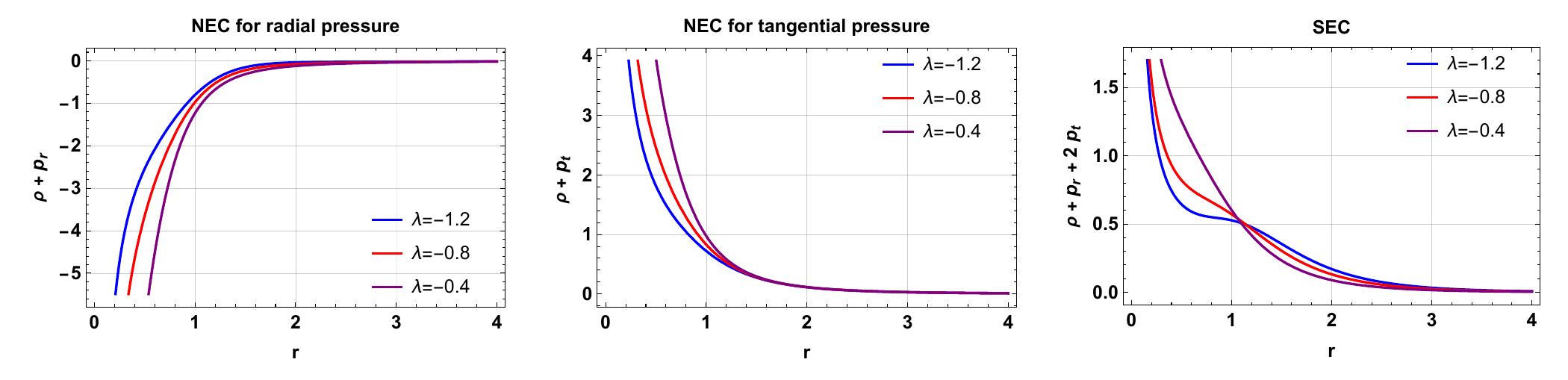}
\caption{Profile shows the behavior of NEC and SEC versus radial coordinate `$r$ ' for different values of model parameter $\lambda$ under NFW-II model. Also, we consider $\rho_0=0.2$, $r_s=1.4$, and $r_0=1.5$.}\label{9}
\end{figure*}

In this subsection, the NFW Model-II energy density profile, which we previously established in Eq. \eqref{4g}, will be the subject of investigation. The shape function from Eq. \eqref{5e} will be incorporated into Equations \eqref{5c} and \eqref{5d} to analyse the Null Energy Condition for both radial and tangential pressures at the throat $r = r_0$. 
\begin{multline}\label{5j}
\rho+p_r\bigg\vert_{r=r_0}= \mathcal{M}_8 r_0 -\frac{\text{\text{e}} \left(4 r_0 ^4 \lambda +\left(1+\text{\text{e}}^2\right) r_0 ^3-4 \text{\text{e}} \lambda \right)}{2 \left(\text{\text{e}} r_0 ^3-2 \lambda \right)^2}\,,
\end{multline}
\begin{multline}\label{5k}
 \rho+p_t\bigg\vert_{r=r_0}=\mathcal{M}_8 r_0 +\frac{\text{\text{e}} \left(3+\text{\text{e}}^2\right) r_0 ^3-4 \left(1+\text{\text{e}}^2\right) \lambda }{4 \left(\text{\text{e}} r_0 ^3-2 \lambda \right)^2}\,.
\end{multline}
Additionally, Fig. \ref{9} provides a visual representation of the energy conditions.\\
Now, we discuss the graphical representation of the energy conditions using the particular form of shape function \eqref{5e}. So, we have presented graphical representations in Figs. \ref{7}-\ref{9} to visualize the energy conditions by varying the coupling parameter $\lambda$. While studying these, we have fixed some parameters like $\rho_0=0.2$ and $r_s=1.4$. Further, we assume the wormhole throat as $r_0=1.5$. The figures \ref{7}-\ref{9} provide insights into the  NEC for radial pressure, NEC for tangential pressure, and SEC for three distinct profiles: the URC model, NFW Model-I, and NFW Model-II. When examining the NEC for radial pressure across different values of $\lambda$, we observe a consistently negative trend, indicating a violation of the NEC across these parameter variations. On the other hand, the NEC for tangential pressure exhibits positive behavior. This intriguing observation suggests that the specific model of $f(R, L_m)$ gravity possesses unique characteristics that enable a positive contribution to the combined energy density and tangential pressure. Consequently, the NEC is violated for all profiles, suggesting that exotic matter may indeed support wormhole solutions within the framework of curvature-matter coupling $f(R,L_m)$ gravity, similar to GR. Notably, the SEC consistently demonstrates a positive behavior across various $\lambda$ values.
\section{Conclusions}\label{sec6}
In recent decades, the exploration of wormhole geometry in scientific research has ignited considerable enthusiasm among researchers. This surge of interest has led to findings suggesting that the potential existence of wormholes within the galactic halo region is supported by the density profiles of URC and NFW dark matter. Within this study, we have considered the examination of wormhole geometry influenced by DM galactic halo profiles, specifically, the URC and cold DM halo with NFW model-I and NFW model-II, within the framework of $f(R, Lm)$ gravity. It is a well-established fact that exact solutions are elusive when dealing with arbitrary $f(R,L_m)$  functions. To surmount this challenge, we have opted for the exploration of two distinct models, as $f(R,L_m)=\frac{R}{2}+L_m^\alpha$ and $f(R,L_m)=\frac{R}{2}+(1+\lambda\,R)L_m $. 
 To begin, we consider the $f(R,L_m) = \frac{R}{2} + L_m^\alpha$ model within the context of DM halo profiles. Our exploration of this model led us to derive shape functions by comparing the energy density of DM halo profiles with that of $f(R,L_m)$ gravity. And, to obtain the exact solutions, we have considered some specific values of $\alpha$. Further, we conducted a comprehensive examination of fundamental characteristics, including the assessment of the flare-out condition of the resulting shape functions in an asymptotic background. It is worth emphasizing that the parameter $\alpha$ in this model plays a significant role in analyzing the shape of the wormholes. Our observations revealed that the flaring out condition is satisfied to the wormhole throat. Furthermore, we conducted an assessment of the NEC within the framework of DM galactic halo models in the vicinity of the throat. Notably, we observed a violation of NEC at the throat, which corresponded to specific parameter values, namely $ \rho_0 = 0.0196$, $r_s = 6$, and $r_0 = 1$ with model parameter $\alpha=2,3$. It is well known that in ordinary GR existance of wormhole throat would have to violate the null energy condition. This can be proven by fallowing the argument given in \cite{Vis} that is if we take a set of congruent null geodesics when it passes through the throat it has to diverse so from the Raychaudhuri Equation one can show that such a diverging null congruent geodesics has to violate the null energy condition.  Additionally, it was ascertained that the SEC is satisfied at the throat of the wormhole.\\
Furthermore, we looked at the exploration of wormhole solutions under the $f(R, L_m) = \frac{R}{2} + (1 + \lambda R) L_m$ model within the context of DM halo profiles. Our investigation entailed deriving wormhole solutions by employing the prescribed shape function as per Eq. \eqref{5e}. We also conducted a comprehensive analysis of the energy conditions, taking into consideration the influence of URC and NFW dark matter galactic halo profiles on the embedded shape function. We noticed that the NEC is violated for the radial pressure, but it is satisfied for the tangential pressure in the context of the $f(R, L_m)$ profile. Additionally, the SEC remains satisfied across all profiles in the vicinity of the wormhole throat. By setting specific limits and values for the model's parameters, we can illustrate the violation of energy conditions, which could suggest the presence of unusual forms of matter. This unique substance might enable wormholes to exist and traverse through the underlying spacetime. Therefore, within the DM galactic halo profiles, embedded wormhole solutions are physically acceptable.\\
In this work, we have taken into consideration the constant redshift function that is $\Phi(r)=\text{constant}$. In this modified gravity, it might be fascinating to study wormholes with non-constant redshift functions in the near future.

\section*{Data Availability Statement}
There are no new data associated with this article.

\section*{Acknowledgments}
LVJ acknowledges University Grant Commission (UGC), Govt. of India, New Delhi, for awarding JRF (NTA Ref. No.: 191620024300). MT acknowledges University Grants Commission (UGC), New Delhi, India, for awarding National Fellowship for Scheduled Caste Students (UGC-Ref. No.: 201610123801). PKS acknowledges the National Board for Higher Mathematics (NBHM) under the Department of Atomic Energy (DAE), Govt. of India, for financial support to carry out the Research project No.: 02011/3/2022 NBHM(R.P.)/R\&D II/2152 Dt.14.02.2022 and IUCAA, Pune, India for providing support through the visiting Associateship program.

\end{document}